\documentstyle[editedvolume,numreferences]{crckapb}

\begin{opening}
\title{DISTURBING THE BLACK HOLE}

\author{Jacob D. Bekenstein}
\institute{Racah Institute of Physics, Hebrew University of Jerusalem\\
Givat Ram, Jerusalem 91904 ISRAEL}

\end{opening}

\runningtitle{DISTURBING THE BLACK HOLE}

\begin{document}

\begin{abstract}
I describe some examples in support of the conjecture that the horizon
area of a near equilibrium black hole is an adiabatic invariant.  These
include a Schwarzschild black hole perturbed by quasistatic scalar fields
(which may be minimally or nonminimally coupled to curvature), a Kerr black
under the influence of scalar radiation at the superradiance treshold, and a
Reissner--Nordstr\"om black hole absorbing a charge marginally.  These
clarify somewhat the conditions under which the conjecture would be true. 
The desired ``adiabatic theorem'' provides an important motivation for a
scheme for black hole quantization.
\end{abstract}

\section{Introduction}

Does the event horizon area of a black hole always grow under external
perturbations ?  Hawking's area theorem \cite{Hawking} would suggest
an affirmative answer whenever classical fields obeying the weak energy
condition are involved.  Nevertheless, one can categorize a variety of
situations in which an external perturbation transmitted through common
fields is slowly applied and relaxed, and does {\it not\/} lead to area
increase.  This classical ``adiabatic invariance'' of horizon area, not yet a
theorem but a collection of examples, is obviously consistent with the
entropy character of black hole area \cite{Bekentropy,HawkingCMP} because in
classical thermodynamics entropy is invariant under slow changes of an
insulated system in thermodynamic equilibrium.  It is also an important
motivation in an approach to black hole quantization
\cite{BekNC,Mukhanov,BekMukhanov,BekBrazil} which has received increasing
attention in the last couple of years. \cite{list}  Keeping the ultimate
application of the adiabatic property to black hole quantization in the back
of the mind will help in grasping the significance of the medley
of examples here garnered.  In collecting these I have had in mind
generating interest in turning the observation into a precise theorem.

Consider a small patch of event horizon area $\delta A$; it is formed by null
generators whose tangents are $l^\alpha=dx^\alpha/d\lambda$, where $\lambda$
is an affine parameter along the generators.  By definition of the 
convergence $\rho$ of the generators, $\delta A$ changes at a rate
\begin{equation}
d\delta A/d\lambda = -2\rho\delta A. 
\label{changeA}
\end{equation}
Now $\rho$ itself changes at a rate given by the optical analogue of the
Raychaudhuri equation (with Einstein's equations already incorporated; I use
units such that $G=c=1$)
\cite{NP,Pirani}
\begin{equation}
d \rho/d\lambda = \rho^2 + |\sigma|^2 + 4\pi
T_{\alpha\beta}\,l^\alpha l^\beta, 
\label{changerho}
\end{equation}
where $\sigma$ is the shear of the generators and $T_{\alpha\beta}$ the 
energy momentum tensor.   The shear evolves according to
\begin{equation}
d  \sigma/d\lambda = 2\rho\sigma + (3\epsilon
-\overline\epsilon)\sigma +C_{\alpha\beta\gamma\delta}\, l^\alpha m^\beta
l^\gamma m^\delta, 
\label{changesigma}
\end{equation}
where $C_{\alpha\beta\gamma\delta}$ is the Weyl conformal tensor,
$m^\alpha$  one of the Newman--Penrose tetrad legs which lies in the
horizon, and $\epsilon$ a pure imaginary parameter.

Many types of classical matter obey the weak energy condition
\begin{equation}
T_{\alpha\beta}\,l^\alpha l^\beta \geq 0. 
\label{energycondition}
\end{equation}
Whenever this is true, $\rho$ can - according to Eq.~(\ref{changerho}) - only
{\it grow\/} or remain unchanged along the generators.  Now were $\rho$ to
become positive at any event along a generator of our horizon patch, then by
Eq.~(\ref{changerho}) it would remain positive henceforth, and indeed grow
bigger.  The joint solution of Eqs.~(\ref{changeA})-(\ref{changerho}) shows
that $\delta A$ would shrink to nought in a finite span of
$\lambda$.\cite{Hawking,MTW}  This vanishing with its implied extinction of
generators would constitute a singularity on the horizon. But it is an axiom
of the subject \cite{Hawking} that event horizon generators cannot end in the
future.  The only way out is to accept that
$\rho\leq 0$ everywhere along the generators, which by Eq.~(\ref{changeA})
signifies that the horizon patch's area can never decrease.  This is the
essence of Hawking's area theorem.

It will be noticed that to keep the horizon area constant requires
$\rho=0$ which by Eqs.~(\ref{changerho})-(\ref{changesigma}) implies that
both $C_{\alpha\beta\gamma\delta}\, l^\alpha m^\beta l^\gamma 
 m^\delta $  and $T_{\alpha\beta}\,l^\alpha l^\beta $ vanish at
the horizon. Vanishing of $C_{\alpha\beta\gamma\delta}\, l^\alpha m^\beta
l^\gamma  m^\delta $ requires that the geometry be quasistationary
to prevent gravitational waves, which are quantified by
$C_{\alpha\beta\gamma\delta}$, from impinging on the horizon.  Thus with a
quasistationary geometry, preservation of the horizon's area requires
\begin{equation}
T_{\alpha\beta}\,l^\alpha l^\beta  = 0 \quad {\rm on\ the\ horizon}.
\label{zerocondition}
\end{equation}

Contrary to the folklore which considers increase of horizon area to be an
almost compulsory consequence of changes in the black hole, I shall here
exhibit a variety of situations for which the conditions that
keep horizon area unchanged occur naturally.  The rule that seems to emerge
is that quasistationary changes of the black hole occasioned by an external
influence will leave the horizon area unchanged.  This means an
``adiabatic theorem'' for black holes must exist.

\section{Black Hole Disturbed by Scalar Charges}

Consider a Schwarzschild black hole with exterior metric
\begin{equation}
ds^2 = - (1-2M/r) dt^2 + (1-2M/r)^{-1} dr^2 + r^2 (d\theta^2 + \sin^2\theta
\, d\varphi^2).
\label{Schwarzschild}
\end{equation}
Suppose sources of a {\it minimally coupled\/} scalar field $\Phi$ are
brought up slowly from infinity to a finite distance of the hole, and then
withdrawn equally slowly. Does this changing influence increase the horizon's
area ?  Given that the changes are quasistatic, the question is just whether
condition (\ref{zerocondition}) is satisfied for all time.  As we shall point
out in Sec.~3, a potential barrier at $r\sim 3M$ screens out quasistatic
scalar perturbations.  For this reason our analysis becomes more than
academic only when the sources actually enter the region $2M<r<3M$.  

If the scalar's sources are weak, one may regard $\Phi$ as a quantity of
first order, and proceed by perturbation theory.  The scalar's
energy--momentum tensor,
\begin{equation}
	T_\alpha{}^\beta  = \nabla_\alpha \Phi \nabla^\beta \Phi - {1\over
2}\,\delta^\beta_\alpha\, \nabla_\gamma\Phi\,\nabla^\gamma \Phi,
\label{energytensor}
\end{equation}
will be of second order of smallness.  I shall suppose the same is true of
the energy--momentum tensor of the sources themselves.  Thus to first order
the metric (\ref{Schwarzschild}) is unchanged.  Neglecting for the moment
time derivatives, the scalar equation outside the scalar's sources can be
written in the form
\begin{equation}
\partial/\partial r \left[ (r^2-2Mr) \partial\Phi/\partial
r\right] - \hat L^2\,\Phi = 0,
\label{scalarequation}
\end{equation}
where  $\hat L^2$ is the usual squared angular momentum operator (but
without the $\hbar^2$ factor).  This equation suggests  looking for a
solution of the form
\begin{equation}
\Phi = \Re \sum_{\ell=0}^\infty\, \sum_{m=-\ell}^\ell 
f_{\ell m}(r)\, Y_{\ell m}(\theta,\varphi),
\label{Phi}
\end{equation} 
where the $Y_{\ell m}$ are the familiar spherical harmonic
(complex) functions.
 Since the 
$Y_{\ell m}$ form a complete set in angular space, any function
$\Phi(r,\theta,\varphi)$ can be so expressed.  And since
$\hat L^2 Y_{\ell m} = \ell(\ell+1) Y_{\ell m}$, the radial and angular
variables separate, and one finds for $f_{\ell m}$ the equation
\begin{equation}
d/d r \left[ (r^2-2Mr) df_{\ell m}/dr\right] - \ell(\ell+1)f_\ell  = 0.
\label{scalarequation2}
\end{equation}
Since the index $m$ does not figure in this equation, I write just
plain $f_\ell(r)$; one may obviously pick  $f_\ell(r)$ to be real.

Let us change from variable $r$ to $x\equiv r/M-1$ and define $F_\ell(x)
\equiv f_\ell(r)$, so that Eq.~(\ref{scalarequation2}) becomes
\begin{equation}
d/dx\left[(1-x^2) F_\ell\right] + \ell(\ell+1)F_\ell = 0.
\label{Fequation}
\end{equation}
This is the Legendre equation of order $\ell$.  Its solutions regular at
the singular point $x=1$ of the equation are the well known Legendre
polynomials
$P_\ell(x)$. Independent solutions are furnished by the Legendre
associated functions $Q_\ell$ which have the general form \cite{MathewsWalker}
\begin{equation}
Q_\ell(x) = {1\over 2} \ln\left[{x+1\over x-1}\right] P_\ell(x)  + {\rm
polynomial\ of\ order\ } (\ell-1) {\ \rm in\ } x. 
\label{Legendrefunctions}
\end{equation}
The associated solutions are thus singular at the horizon $x=1$, and 
must not be included in $\Phi$, as the following argument makes clear.

We obviously require that the event horizon remain regular under the scalar's
perturbation; otherwise the black hole is destroyed and our discussion is
over before it began.  A minimal requirement for regularity is that
physical invariants like $\Upsilon_1\equiv T_\alpha{}^\alpha$, 
$\Upsilon_2\equiv T_\alpha{}^\beta T_\beta{}^\alpha$, 
$\Upsilon_3\equiv T_\alpha{}^\beta T_\beta{}^\gamma T_\gamma{}^\alpha$, {\it
etc.,\/} be bounded, for divergence of any of them would surely induce
curvature singularities via the Einstein equations.  By
Eq.~(\ref{energytensor}) the invariant $\Upsilon_k$ is always proportional to
$(\Phi,_\alpha\Phi,^\alpha)^k$.  Now to lowest order of smallness the 
metric components that enter into $\Upsilon_k$ are just the Schwarzschild 
ones.  In particular, in our static case $\Phi,_\alpha\Phi,^\alpha =
(1-2M/r)\Phi,_r^2 + \cdots\ $. One must thus require
\begin{equation}
(1-2M/r)^{1/2}\, \Phi,_r {\ \rm bounded\ at\ horizon.}
\end{equation}
Since this condition must hold for every $\theta$ and $\varphi$, it follows
from  the independence of the various $Y_{\ell m}$ in Eq.~(\ref{Phi}) that
\begin{equation}
\forall \ell: \quad (1-2M/r)^{1/2}\, df_\ell/dr {\rm\ \ bounded}.
\label{Tmunu}
\end{equation}
But according to Eq.~(\ref{Legendrefunctions}) the $Q_\ell$ are too
singular to satisfy this equation, {\it i.e.,\/}  $\sqrt{x-1}\, Q'_\ell(x)
\rightarrow \infty$ as $x\rightarrow 1$.  Thus one must discard the
$Q_\ell$ from the set of radial solutions relevant in the region between the
inermost source and the horizon.

Thus in this inner region of the black hole exterior we have, to first order
in perturbation theory, the exact solution  
\begin{equation}
\Phi = \Re \left(\sum_{\ell=0}^\infty\, \sum_{m=-\ell}^\ell 
C_{\ell m}\, P_\ell(r)\, Y_{\ell m}(\theta,\varphi)\right),
\label{newPhi}
\end{equation}
where the (complex) coefficients $C_{\ell m}$ permit us to match the solution
to every distribution of sources by the usual methods.  As those sources are
moved around slowly, the $C_{\ell m}$ will change slowly (we shall
investigate the question of changes at finite speed in Sec.~3). 
Now according to Eq.~(\ref{energytensor}), 
\begin{equation}
 T_r{}^r - T_t{}^t = (1-2M/r)\, \Phi,_r^2.
\label{difference}
\end{equation}
Because $\Phi$ is completely regular down to the horizon, this shows that
\begin{equation}
\lim_{r\rightarrow 2M}\ (T_r{}^r - T_t{}^r) =  0.
\label{Trr-Ttt}
\end{equation}
We now explain the significance of this general result \cite{AGK,BekMayo} for
our specific problem. 

Any 3D--hypersurface of the form $\{\forall t, r={\rm const.}\}$ has a
tangent $\tau^\alpha = \delta_t{}^\alpha$ with norm $-(1-2M/r)$ as well as
the normal $\eta_\alpha =
\partial_\alpha (r-{\rm const.}) =\delta_\alpha{}^r$ with norm $(1-2M/r)$. 
The vector $N^\alpha\equiv \tau^\alpha+(1-2M/r) \eta^\alpha$ is obviously
null, and as $r\rightarrow 2M$, both its covariant and contravariant forms
remain well defined.  Indeed, as $r\rightarrow 2M$ the other two vectors
become null; in contrast with them, $N^\alpha$ remains well behaved at the
horizon, so that it must there be proportional (with finite nonvanishing
proportionality constant) to $l^\alpha$, the tangent to the horizon
generator.   This can be verified by remarking that $N^\alpha$, just as
$l^\alpha$, is future pointed ($N^t>0$) as well as outgoing ($N^r>0$).  Now
at any point
$r\ge 2M$
\begin{equation}
 T_{\alpha\beta} N^\alpha N^\beta =   
T_t{}^t\, N_t\,N^t + T_r{}^r\, N_r\,N^r= (T_r{}^r - T_t{}^r)\, N_r\,N^r.
\label{TNN}
\end{equation} 
But $ N_rN^r \rightarrow 0$ at the horizon, so that in view of
Eq.~(\ref{Trr-Ttt}), $T_{\alpha\beta} l^\alpha l^\beta = 0$.

Since Eq.~(\ref{zerocondition}) is satisfied, the horizon's area does not
change under the action of the scalar field.  This conclusion is obviously
conditional on the changes of the scalar's sources taking place sufficiently
slowly, for otherwise Eq.~(\ref{scalarequation}) would be invalid, and the
components of $T_\mu{}^\nu$ I employed would be affected.  Our result
supports the notion that adiabatic perturbations of a Schwarzschild black
hole do not change the area, so that horizon area is an adiabatic invariant.

Notice that $T_{\alpha\beta} l^\alpha l^\beta$ vanishes because it
is a product of {\it two\/} vanishing factors.  This suggests that the area
invariance result will remain valid under small ``perturbations'' to
our scenario.   Let us investigate the issue of time dependence of the scalar
field; it is important because I am contemplating moving the sources of the
scalar field so that it is never perfectly static as assumed heretofore.

\section{The Time Dependent Problem}

Let us retain in the scalar equation the time derivatives:
\begin{equation}
-{r^4\over (r^2-2Mr)}{\partial^2\Phi \over \partial t^2} +
{\partial\over \partial r} \left[ (r^2-2M) {\partial\Phi\over\partial r}
\right] - \hat L^2\,\Phi = 0.
\label{tdependence}
\end{equation}
In analogy with Eq.~(\ref{newPhi})  we now look for a solution of the form
\begin{equation}
\Phi = \Re \int_{0}^{\infty} d\omega\sum_{\ell=0}^\infty\,
\sum_{m=-\ell}^\ell C_{\ell m}(\omega)\, f_{\ell}(\omega,r)\, Y_{\ell
m}(\theta,\varphi) e^{-\imath\omega t}.
\label{Phitd}
\end{equation}
In terms of Wheeler's ``tortoise'' coordinate $r^* \equiv r + 2M\ln(r/2M-1)$,
for which the horizon resides at $r^*=-\infty$,  the equation satisfied by
the new radial function $H_{\ell\omega}(r^*)\equiv rf_\ell(\omega,r)$ is
\cite{MTW}
\begin{equation}
-{d^2H_{\ell\omega}\over dr^{*2}} + \left(1-{2M\over r}\right)\left({2M\over
r^3} + {\ell(\ell+1)\over r^2}\right) H_{\ell\omega}= \omega^2
H_{\ell\omega}. 
\label{Feq}
\end{equation}

The analogy between Eq.~(\ref{Feq}) and the Schr\"odinger eigenvalue
equation permits the following analysis \cite{MTW} of the effects of distant
scalar sources on the black hole horizon.  Waves with ``energy'' $\omega^2$
on their way in from a distant source run into a positive potential, the
product of the two parentheses in Eq.~(\ref{Feq}).  The potential's peak is
situated at
$r\approx 3M$ for all $\ell$.  Its  height is $0.0264 M^{-2}$ for $\ell=0$,
$0.0993 M^{-2}$ for $\ell=1$ and $\approx 0.038\,\ell(\ell+1) M^{-2}$ for
$\ell\ge 2$. Therefore, waves with any $\ell$ and $\omega < 0.163 M^{-1}$
coming from sources at $r\gg 3M$ have to tunnel through the potential barrier
to get near the horizon.  As a consequence, the wave amplitudes that
penetrate to the horizon are small fractions of the initial amplitudes, most
of the waves being reflected back.  In fact, the tunnelling coefficient
vanishes in the limit $\omega\rightarrow 0$. \cite{MTW} This means that
adiabatic perturbations by distant sources (which surely means they only
contain Fourier components with $\omega\ll M^{-1}$) perturb the horizon very
weakly (this is just an inverse of Price's theorem \cite{MTW} that a totally
collapsed star's asymptotic geometry preserves no memories of the star's
shape).  Thus one would not expect significant growth of horizon area from
scalar perturbations originating in distant sources.

What if the scalar's sources are moved into the region $2M < r < 3M$ inside
the barrier ?  They will now be able to perturb the horizon; do they change
it's area ?  To check I look for the solutions of Eq.~(\ref{Feq}) in the
region near the horizon where the potential is small compared to
$\omega^2$; according to the theory of linear second order differential
equations they are of the form
\begin{equation}
H_{\ell\omega}(r^*) = \exp(\pm \imath\omega r^*)\times[1 +  O(1-2m/r)].
\label{solution}
\end{equation}
The Matzner boundary condition \cite{Matzner} that the physical solution be
an ingoing wave as appropriate to the absorbing character of the horizon
selects the sign in the exponent as negative.  Hence the typical term in
$\Phi$ is
\begin{equation}
{1 +  O(1-2m/r)\over r} P_\ell(\cos\theta)\,\cos\psi;\qquad \psi\equiv
\omega(r^*+t)-m\varphi.
\label{term}
\end{equation}  

Is the perturbation well behaved at the horizon ?  In particular, are all
the invariants $\Upsilon_k$ of Sec.~2 (or equivalently
$\Phi,_\alpha\Phi,^\alpha$) bounded there ?  Let us first look at a $\Phi$
composed of a single mode like that in Eq.~(\ref{term}).  An explicit
calculation on the Schwarzschild background using $dr^*/dr = (1-2M/r)^{-1}$
gives, after a miraculous cancellation of terms divergent at the horizon
(pointed out by A. Mayo),
\begin{equation}
\Phi,_\alpha\Phi,^\alpha \propto
{m^2 P_\ell{}^2\sin^2\psi\over r^4\sin^2 \theta}
 +\left({dP_\ell\over d\theta}\right)^2{\cos^2\psi\over r^4}
+{\omega \sin(2\psi)\over r^3}P_\ell^2 + \cdots\, ,
\label{miracle}
\end{equation}
where ``$\ \cdots\ $'' here and henceforth denote terms that vanish as
$r\rightarrow 2M$.  This expression is bounded at the horizon.    Now suppose
$\Phi$ is the sum of two modes like (\ref{term}).  Let us label the various
parameters with subscripts ``1'' and ``2''.  Then a calculation gives
$\Phi,_\alpha\Phi,^\alpha$ as consisting of three groups of terms, two of
them of form (\ref{miracle}) with subscripts 1 and 2, respectively, and a
third of the form
\begin{eqnarray}
{m_1 m_2 P_{\ell_1}P_{\ell_2}\sin\psi_1\sin\psi_2\over
r^4\sin^2 \theta}
 +\left({dP_{\ell_1}\over d\theta}\right)\left({dP_{\ell_2}\over
d\theta}\right){\cos\psi_1\cos\psi_2 \over r^4} +
\nonumber \\
+{\omega_1 \sin\psi_1\cos\psi_2 +\omega_2 \sin\psi_2\cos\psi_1\over
r^3}P_{\ell_1}P_{\ell_2} + \cdots
\end{eqnarray}
This is also bounded. By induction any $\Phi$ of form (\ref{Phitd}) will
give a bounded $\Phi,_\alpha\Phi,^\alpha$. Thus all the $\Upsilon_k$ are
bounded at $r=2M$, and a generic scalar perturbation does not disturb the
horizon unduly.

The extent by which the {\it shape\/} of the horizon is perturbed must be
linear in the magnitude of the invariant $\Upsilon_1$ (Einstein's equations
have $T_{\alpha\beta}$ as source, not $T_\alpha{}^\gamma T_\gamma{}^\beta$).
It is then clear from both our results that this perturbation is of order
$O(\omega^0)$ generically, and of $O(\omega)$ in the monopole case.
I now show that the change in the horizon area is of $O(\omega^2)$, so
that for small $\omega$ the area is (relatively) invariant.

Because now there is time variation and $T_t{}^r\neq 0$, Eq.~(\ref{TNN}) has
to be generalized to the form
\begin{equation}
T_{\alpha\beta} N^\alpha N^\beta = (T_r{}^r - T_t{}^r)\, N_r\,N^r + 2
T_t{}^r N_r N^t.
\label{TNNnew}
\end{equation}
From $N^\alpha$'s definition in Sec.~2 we have $N_rN^r = 1-2M/r$ and  $
N_r N^t=1$. And from Eq.~(\ref{energytensor}) it is clear that $T_r{}^r -
T_t{}^t=
\Phi,_r\Phi,^r-\Phi,_t\Phi,^t$ while $T_t{}^r = \Phi,_t\Phi,^r$.  Thus
\begin{equation}
T_{\alpha\beta} N^\alpha N^\beta = [\Phi,_t + (1-2M/r)\Phi,_r]^2.
\end{equation}
If one now substitutes a $\Phi$ made up of a single mode like in
Eq.~(\ref{term}),  one concludes that
\begin{equation}
T_{\alpha\beta} l^\alpha l^\beta \propto {\omega^2 P_\ell^2\sin^2\psi\over
r^2} + \cdots\, .
\label{term2} 
\end{equation}
A quick way to this result is to recognize that
$l^\alpha\propto \tau^\alpha\equiv (\partial/\partial t)^\alpha$ because the
horizon generators must lie along the only Killing vector of the problem. 
In view of Eq.~(\ref{energytensor}) and the null character of $l^\alpha$,
\begin{equation}
T_{\alpha\beta} l^\alpha l^\beta \propto (\Phi,_\alpha \tau^\alpha)^2 =
(\partial\Phi/\partial t)^2, 
\end{equation}
which reproduces Eq.~(\ref{term2}). And if one substitutes the generic $\Phi$,
the proportionality to the square of frequency will obviously remain.

Thus, when scalar field sources are moved inside the barrier, they perturb
the geometry, and the horizon's shape in particular, by an amount which does
not, in general, vanish as the perturbations changes very slowly.  By
contrast, the rate of change of the horizon area vanishes as the square of
the typical Fourier frequency of the perturbation.  In this sense the horizon
area is an adiabatic invariant. 

\section{Generalization to Nonminimally Coupled Field}

To what extent is our result generic ?  For example, does it depend on the
nature of the interaction ? One can probe this question by replacing our
minimally coupled scalar field by one coupled nonminimally to curvature.
The scalar equation is replaced by \cite{Penrose}
\begin{equation}
(\nabla_\alpha\nabla^\alpha - \xi R) \Phi = 0,
\label{confscalarequation}
\end{equation}
where $R$ denotes the Ricci scalar and $\xi\neq 0$ is a real constant
measuring the extent of nonminimal coupling; $\xi=1/6$ corresponds to a
conformally invariant scalar equation.  The corresponding energy--momentum
tensor is \cite{Parker}
\begin{equation}
T_\alpha{}^\beta = \nabla_\alpha \Phi \nabla^\beta \Phi - {1\over 2}
\nabla_\alpha \Phi \,\nabla^\alpha \Phi\,\delta^\beta_\alpha	 - \xi
(\nabla_\alpha \nabla^ \beta \Phi^2 -  \delta^\beta_\alpha \nabla_\mu
\nabla^\gamma \Phi^2  -  G_\alpha{}^\beta \Phi^2),	      
\end{equation}
where $G_\alpha{}^\beta$ denotes the Einstein tensor.  Evaluating it with
help of the Einstein equations one obtains, in the region outside the
scalar's sources where they do not contribute to $	T_\alpha{}^\beta$,
\begin{equation}
	T_\alpha{}^\beta  =
{\nabla_\alpha \Phi \nabla^\beta \Phi - {1\over 2}\,\delta^\beta_\alpha\,
\nabla_\gamma \Phi\,\nabla^\gamma \Phi	 - \xi (\nabla_\alpha
\nabla^\beta \Phi^2 - \delta^\beta_\alpha \nabla_\gamma \nabla^\gamma
\Phi^2) \over 1-8\pi G \xi\Phi^2}.	      
\label{confscalarTmunu}
\end{equation}

The following analysis is carried out with neglect of time variation as in
Sec.~2.  If again one regards $\Phi$ as of first order of smallness, then it
is clear that $T_\alpha{}^\beta$ is of second order and consequently the
lowest correction to the Schwarzschild metric Eq.~(\ref{Schwarzschild}) is of
second order.  Likewise, $R$ in Eq.~(\ref{confscalarequation}) is of second
order.  Therefore, in the static case one may again get $\Phi$ to first order
by solving Eq.~(\ref{scalarequation}), and $\Phi$ is given by the series in
Eq.~(\ref{Phi}) with radial functions which are superpositions of Legendre
polynomials $P_\ell$ and Legendre functions $Q_\ell$.  

To select out the physical combinations of these, I again require - the
argument is exactly like in Sec.~2 - that every {\it diagonal\/} component of
$T_\alpha{}^\beta$ be bounded.  In particular one has from
Eq.~(\ref{confscalarTmunu}) that
\begin{equation}
	T_t^t-T_\varphi^\varphi =
	{2\xi(1-3M/r)\Phi\,\Phi,_r \over  r(1-8\pi G\xi \Phi^2)}\quad {\rm bounded\
at\ horizon}.
\label{Ttt-Tphph}
\end{equation}
It is clear from this that $\Phi$ itself must be bounded at $r=2M$, for if it
were to diverge there, then $|T_t^t-T_\varphi^\varphi| \sim |d\ln\Phi/dr|
\rightarrow\infty$.  I thus  discard all the $Q_\ell$ functions from $\Phi$,
thereby returning to the form (\ref{newPhi}): $\Phi$ is regular at the
horizon. 

Now
\begin{equation}
 T_t{}^t-T_r{}^r = \left(1-{2M\over r}\right){ (2\xi-1) \Phi,_r^2 +
2\xi\,\Phi\,\Phi,_{rr}
\over 1-8\pi G \xi \Phi^2}.
\label{Ttt-Trr} 
\end{equation}
From this it is clear that $T_t{}^t-T_r{}^r \rightarrow 0$ at $r=2M$ unless
$8\pi G \xi \Phi^2\rightarrow 1$ there.  This last possibility must be
excluded for it would be equivalent to having $\ln(1-8\pi G \xi
\Phi^2)\rightarrow -\infty$, whereby $d\ln(1-8\pi G \xi
\Phi^2)/dr$ would necessarily diverge at $r=2M$.   But this derivative
occurs as a factor in $T_t^t-T_\varphi^\varphi$ and its divergence would not
be compensated for; thus Eq.~(\ref{Ttt-Tphph}) would be contradicted.  Since
$T_t{}^t-T_r{}^r \rightarrow 0$ on the horizon, the arguments at the end of
Sec.~2 can be repeated to show that the area of the black hole is unchanged
by the scalar perturbation. Although we shall not go here into the time
dependent problem, it seems that coupling the field nonminimally makes little
difference regarding the adiabatic invariance of the horizon.

A. Mayo \cite{Mayo} has worked out the effect of static electromagnetic 
perturbations on the horizon and shown that they also leave its area
unchanged. 

\section{Waves Impinging on a Kerr Black Hole}

In the above two examples when the perturbation is removed, the black hole
exterior must return to exact Schwarzschild form because a static spherical
black hole with no electric charge cannot retain scalar hair
\cite{noscalarhair}.  Further, the black hole returns to its
original mass since, for the Schwarzschild case, the horizon area and the
mass are related one--to--one, and the area has not changed.  The question
thus arises, does adiabatic invariance of the area continue to hold when the
black hole's perturbation is accompanied by a net change in some of the
Wheeler black hole parameters  mass, charge and angular momentum ?   We now
show the answer is affirmative for adiabatic scalar perturbations of the Kerr
black hole.

In the Kerr case the meaning of ``adiabatic'' needs to be refined.  It is
well known that a static ($\omega=0$) but nonaxisymmetric perturbation of a
Kerr black hole, such as would be caused by field sources held in its
vicinity at rest with respect to infinity, necessarily causes an increase in
horizon area. \cite{HawkingHartle}   However, static perturbations in this
sense are not adiabatic from the local point of view.  Because of the
dragging of inertial frames, \cite{MTW} any nonaxisymmetric static field is
perceived by momentarily radially stationary local inertial observers as
endowed with temporal variation as these observers are necessarily dragged
through the field's spatial inhomgeneity.  At the horizon the dragging
frequency is the hole's rotational frequency $\Omega$, and a field component
with azimuthal ``quantum'' number $m$ is seen to vary with temporal frequency
$m\Omega$ which need not be small.  Evidently, ``adiabatic'' must here
mean that {\it according to momentarily radially stationary local inertial
observers\/}, the perturbation has only low frequency Fourier components.

To see these concepts in practice, consider a Kerr black hole of mass $M$ and
angular momentum $J$.  I shall not need the metric; all that is important
here is that the horizon area is
\begin{equation}
A = 4\pi\left[\left(M+\sqrt{M^2-(J/M)^2}\right)^2+(J/M)^2\right].
\label{Kerrarea}
\end{equation}
Let sources distributed well away from the hole radiate upon it a {\it
weak\/} scalar wave of the form
\begin{equation}
\Phi = f(r)\, Y_{\ell\,m}(\theta,\varphi)\,e^{-\imath\omega t}.
\label{wave} \end{equation}
In the spirit of perturbation theory I shall neglect the gravitational
waves so produced.  The black hole geometry will eventually be changed by
interaction with this wave, but since the latter is taken to be weak, I shall
assume that the change amounts to a transition from one Kerr geometry to
another with slightly different $M$ and
$J$.  In the final analysis such assumption is justified by the stability of
the Kerr geometry. Since the geometry thus remains axisymmetric and
stationary after the change, the wave preserves its form (\ref{wave}) over
all time.  Long ago Starobinskii \cite{Starobinskii} showed that for small
$\omega-m\Omega$ the absorption coefficient for a wave like (\ref{wave}) has
the form
\begin{equation}
\Gamma = K_{\omega\ell\,m}(M, J)\cdot \left(\omega -\Omega\, m\right),
\label{Gamma}
\end{equation}
where
\begin{equation}
\Omega\equiv {J/M\over {r_{\cal H}}^2  + (J/M)^2};\quad {\rm with}\quad 
r_{\cal H}
\equiv M+\sqrt{M^2-(J/M)^2}
\label{Omega}
\end{equation}
is the rotational angular frequency of the hole, while
$K_{\omega\ell\,m} (M, J)$ is a positive coefficient.  If one chooses
$\omega=m\Omega$ (static perturbation in the eyes of the local inertial
observers), the wave is perfectly reflected and no change of the black hole
parameters ensues. 

By choosing $\ \omega-\Omega\, m\ $ slightly positive, one arrange for a
small fraction of the wave to get absorbed.  Now imagine surrounding the
system by a large spherical mirror.  The part of the wave reflected off the
black hole gets reflected perfectly back towards it by the mirror. The wave
thus bounces back and forth between the two and a sizable fraction of its
energy and angular momentum will eventually get absorbed by the hole. 
Similarly, if one makes $\omega - m\Omega$ slightly negative, the wave gets
amplified upon reflection off the hole (Zel'dovich--Misner superradiance
\cite{Zeldovich}) and the repeated reflections make it stronger than
originally. Since both processes envisioned take place over many cycles of
reflection (a long interval of $t$ time), the consequent (substantial)
changes of $M$ and $J$ occur {\it adiabatically\/} according to distant
observers.  This is addition to the adiabatic nature of the perturbing wave
as seen by local observers. 

A simple calculation shows that small changes of the horizon area are
related to those of $M$ and $J$ by \cite{Bekthesis,Bekentropy,BCH}
\begin{equation}
\Delta A = \left(\Delta M - \Omega \Delta J\right)/\Theta_{K},
\label{dAreaK}
\end{equation}
where
\begin{equation}
\Theta_{K} \equiv {1\over 2} A^{-1} \sqrt{M^2-(J/M)^2} 
\label{coeff}
\end{equation}
is the surface tension \cite{Bekthesis} or surface gravity \cite{BCH}  of the
black hole. The overall changes $\Delta M$ and $\Delta J$ must stand in the
ratio $\omega/m$.  This can be worked out from the energy--momentum tensor,
but is immediately clear if one thinks of the wave as composed of quanta,
each  with energy $\hbar\omega$ and angular momentum $\hbar m$, and using
conservation energy and angular momentum.  Since $\omega\approx \Omega m$, it
is seen from Eq.~(\ref{dAreaK}) that if the black hole is not extremal
($J\neq M^2$ and so $\Theta_K\neq 0$), $\Delta A \approx 0$ to the accuracy
of the former equality.  Therefore, Kerr horizon area is invariant during
adiabatic changes of the mass and angular momentum as judged globally.  These
changes come about from field perturbations which are adiabatic from the
perspective of local inertial observers, as stated earlier.  It is under
these two conditions that the  horizon area is an adiabatic invariant.  

The last conclusion is, however, inapplicable to the extremal Kerr black
hole ($J=M^2$).  In this case $\Theta_K = 0$ so one cannot use
Eq.~(\ref{dAreaK}) to calculate the change in area, but must work directly
with Eq.~(\ref{Kerrarea}).  From Eq.~(\ref{Omega}) one learns that
$\Omega=1/2M$ so that $\Delta J =\Omega^{-1} \Delta M = 2M\Delta M$. 
Replacing $M\rightarrow M+\Delta M$ and $J\rightarrow J+2M\Delta M$ in
Eq.~(\ref{Kerrarea}), and substracting the original expression gives 
\begin{equation}
\Delta A = 8 \pi(2+\surd 2)  M\Delta M + O((\Delta M)^2).
\end{equation}
This is not a small quantity; a generic addition of mass $\Delta M$ will
give a $\Delta A$ of the same order.  Thus horizon area of an extremal Kerr
hole is not an adiabatic invariant.   Sec.~6  gives one more example of
the departure of extremal black horizon area from adiabatic invariance.

For nonextremal black holes the conclusion that horizon area is an adiabatic
invariant extends to perturbations by electromagnetic waves.  One only has
to replace the $Y_{\ell\,m}(\theta,\varphi)$ in the wave by an electric or
magnetic type {\it vector\/} spherical harmonic to describe the
electromagnetic modes.  The conclusion is the same.

\section{Particle Absorption by Reissner--Nordstr\"om Black Hole}

Thus far I have illustrated the adiabatic invariant character of black hole
horizon area under field--black hole interaction.  The present example
focuses rather on point particle--black hole interaction.  It is none
other than the Christodoulou reversible process \cite{Christodoulou} for a
Reissner--Nordstr\"om black hole.  

Consider a Reissner--Nordstr\"om black hole of mass $M$ and positive charge
$Q$. The exterior metric is
\begin{equation}
ds^2 = - \chi\, dt^2 +  \chi^{-1}\, dr^2 + r^2 (d\theta^2 + d\varphi^2),
\label{RNmetric}
\end{equation}
with
\begin{equation}
\chi \equiv 1 - 2M/r + Q^2/r^2.
\label{chi}
\end{equation}
At infinity one shoots in radially a classical point particle of mass $m$
and positive charge $\varepsilon$ with total relativistic energy adjusted to
the value 
\begin{equation}
E=\varepsilon Q/r_{\cal H}.
\label{poised}
\end{equation}
Here $r_{\cal H}$ is the $r$ coordinate of the infinite red shift surface
($\chi = 0$), which by Vishveshwara's theorem
\cite{Vishu} coincides with the event horizon:
\begin{equation}
r_{\cal H} = M + \sqrt{M^2 - Q^2}
\label{r_H}
\end{equation}
In Newtonian terms this particle should marginally reach the horizon where
its potential energy just exhausts the total energy.   The relativistic
equation of motion leads to the same conclusion.

For consider the action for the radial motion,
\begin{equation}
S =  \int{ \left[-m\,\sqrt{\chi\, (dt/d\tau)^2 - (dr/d\tau)^2/\chi}  -
\varepsilon A_t\  dt/d\tau\,\right]\,d\tau},
\label{action}
\end{equation}
where $\tau$, the proper time, acts as a path parameter, and $A_t = Q/r\ $ is
the only nontrivial component of the electromagnetic 4--potential.  The
stationary character of the background metric and field means that there
exists a conserved quantity, namely 
\begin{equation}
E = - \delta S/ \delta (dt/d\tau) =  {m\,\chi\over \sqrt{\chi\, (dt/d\tau)^2
- (dr/d\tau)^2/\chi}}\ {dt\over d\tau} + {\varepsilon\,Q\over r}.
\label{conserved}
\end{equation}
We also know that the norm of the velocity is conserved.  This
together with the definition of proper time gives $\sqrt{\chi\, (dt/d\tau)^2 -
(dr/d\tau)^2/\chi} = 1$.  Substituting $dt/d\tau$ from here in
Eq.~(\ref{conserved}) gives
\begin{equation}
E =  m\,\sqrt{\chi + (dr/d\tau)^2}+ {\varepsilon\,Q\over r}.
\label{newenergy}
\end{equation}
It is easy to see that this is precisely the total energy of the
particle, for at large distances from the hole, $E \approx m+m\upsilon^2/2 -
m\,M/r + \varepsilon\,Q/r$ (sum of rest, kinetic, gravitational and
electrostatic potential energies).  Setting
$E =
\varepsilon\,Q/r_{\cal H}$ shows that the radial motion has a turning point
($dr/d\tau = 0$) precisely at the horizon
$[\chi(r_{\cal H}) = 0]$. 
  
Because the particle's motion has a turning point at the horizon, it
gets accreted by it.  The area of the horizon is originally 
\begin{equation}
A = 4\pi {r_{\cal H}}^2  = 4\pi\left(M+\sqrt{M^2-Q^2}\right)^2,
\label{AreaRN}
\end{equation}
and the (small) change inflicted upon it by the absorption of the particle is
\begin{equation}
\Delta A = (\Delta M - Q\,\Delta Q/r_{\cal H}) /\Theta_{RN}
\label{dAreaRN}
\end{equation}
with
\begin{equation}
\Theta_{RN} \equiv {1\over 2} A^{-1} \sqrt{M^2-Q^2} 
\label{thetaRN}
\end{equation}
being the surface gravity analogous to the previous $\Theta_K$. Thus if the
black hole is not extremal so that $\Theta_{RN}\neq 0$, $\Delta A=0$ because
$\Delta M=E=\varepsilon Q/r_{\cal H}$ while $\Delta Q=\varepsilon$. 
Therefore, the horizon area is invariant under the accretion of the particle
from a turning point.

To a momentarily radially stationary local inertial observer, the particle
in question hardly moves radially as it is  accreted.  Thus its assimilation
is adiabatic.  By contrast, if $E$ were larger than in (\ref{poised}), the
particle would not try to turn around at the horizon, and the local observer
would see it moving radially at finite speed and being assimilated quickly. 
And the horizon's area would increase upon its accretion, as is easy to check
from the previous argument. Thus invariance of the area goes
hand in hand with adiabatic changes at the horizon as judged by
local observers at the horizon.  Needless to say, the changes in $M$ and
$Q$ also occur very slowly as judged by distant observers.  This is the
double sense in which the area is an adiabatic invariant. 

The above conclusions fail for the extremal Reissner--Nordstr\"om
black hole.  When $Q=M$, $\sqrt{M^2-Q^2}$ in Eq.~(\ref{AreaRN}) is unchanged 
to $O(\varepsilon^2)$ during the absorption, so that $\Delta A =8\pi ME$. 
This is not a small change, so the horizon's area is not an adiabatic
invariant.  It is clear, as already noted earlier, that extremal black holes
behave differently from generic black holes in this as in other phenomena. 

\section{Conclusions}

The examples here collated suggest the existence of a theorem which would
state that, classically, under suitably adiabatic changes (two such are here
characterized) of a black hole in equilibrium, the area of its event horizon
does not change.  Aside from harmonizing with the understanding that horizon
area represents entropy, this theorem would provide a formal motivation for
quantizing the black hole in the spirit of the ``old quantum theory'' or
Bohr--Sommerfeld quantization.  The implications of such quantization have
already been considered. \cite{BekNC,Mukhanov}

I thank Avraham Mayo for criticism.  This contribution is based on research
supported by a grant from the Israel Science Foundation.

\end{document}